\documentclass[12pt]{article}

\usepackage{scicite}

\usepackage{times}
\usepackage{amssymb}
\usepackage{subfigure}
\usepackage{graphicx}

\usepackage[style=science,defernumbers=true,backend=bibtex]{biblatex}
\usepackage[final]{microtype}
\usepackage{csquotes,lmodern}
\usepackage{hyperref}
\usepackage{deluxetable}

\hypersetup{hidelinks}

\usepackage[labelfont=bf]{caption}

\renewcommand{\figurename}{Fig.}

\usepackage{filecontents}
\AtEveryBibitem{\clearfield{month}}
\AtEveryCitekey{\clearfield{month}}

\makeatletter

\csnumgdef{blx@entrycount}{0}
\AtEveryBibitem{%
  \csnumgdef{blx@entrycount}{\csuse{blx@entrycount}+1}}

\appto{\newrefsegment}{%
  \csnumgdef{blx@entrycount@\the\c@refsegment}{\csuse{blx@entrycount}+1}}

\defbibcheck{onlynew}{%
  \ifnumless{\thefield{labelnumber}}{\csuse{blx@entrycount@\the\c@refsegment}}
    {\skipentry}
    {}}

\makeatother

\newcommand\ionpat[2]{#1$\;${\scshape{#2}}}

\newenvironment{sciabstract}{%
\begin{quote} \bf}
{\end{quote}}

\renewcommand\refname{References and Notes}

\newcounter{lastnote}

\author{\normalsize Patrick L. Kelly$^{1\ast}$,  Steven A. Rodney$^2$, Tommaso Treu$^{3}$, Ryan J. Foley$^{4}$,\\ 
\normalsize  Gabriel Brammer$^5$, Kasper B. Schmidt$^{6}$, Adi Zitrin$^7$, Alessandro Sonnenfeld$^{3}$, \\
\normalsize  Louis-Gregory Strolger$^{5,8}$, Or Graur$^{9}$, Alexei V. Filippenko$^1$, Saurabh W. Jha$^{10}$, \\ 
\normalsize  Adam G. Riess$^{2,5}$, Marusa Bradac$^{11}$,  Benjamin J. Weiner$^{12}$, Daniel Scolnic$^{2}$, \\
\normalsize Matthew A. Malkan$^{3}$,  Anja von der Linden$^{13}$, Michele Trenti$^{14}$, Jens Hjorth$^{13}$,  \\
\normalsize Raphael Gavazzi$^{15}$, Adriano Fontana$^{16}$, Julian Merten$^{7}$, Curtis McCully$^{6}$, \\
\normalsize Tucker Jones$^{6}$, Marc Postman$^{5}$, Alan Dressler$^{17}$, Brandon Patel$^{10}$, S. Bradley Cenko$^{18}$, \\
\normalsize Melissa L. Graham$^{1}$, and Bradley E. Tucker$^{1}$ \\ 
\footnotesize{$^1$Department of Astronomy, University of California, Berkeley, CA 94720} \\ 
\footnotesize{$^2$Department of Astronomy, The Johns Hopkins University, Baltimore, MD 21218} \\ 
\footnotesize{$^3$Department of Physics and Astronomy, University of California, Los Angeles, CA 90095} \\ 
\footnotesize{$^4$University of Illinois at Urbana-Champaign, 1002 W. Green Street, Urbana, IL 61801} \\ 
\footnotesize{$^5$Space Telescope Science Institute, 3700 San Martin Drive, Baltimore, MD 21218} \\ 
\footnotesize{$^6$Department of Physics, University of California, Santa Barbara, CA 93106} \\ 
\footnotesize{$^7$California Institute of Technology, MC 249-17, Pasadena, CA 91125} \\ 
\footnotesize{$^8$Western Kentucky University, 1906 College Heights Blvd., Bowling Green, KY 42101} \\ 
\footnotesize{$^9$CCPP, New York University, 4 Washington Place, New York, NY 10003} \\ 
\footnotesize{$^{10}$Rutgers, The State University of New Jersey, Piscataway, NJ 08854} \\ 
\footnotesize{$^{11}$Department of Physics, University of California, Davis, CA 95616 } \\ 
\footnotesize{$^{12}$Steward Observatory, University of Arizona, Tucson, AZ 85721 } \\ 
\footnotesize{$^{13}$Dark Cosmology Centre, Juliane Maries Vej 30, 2100 Copenhagen, Demark} \\ 
\footnotesize{$^{14}$School of Physics, University of Melbourne, VIC 3010, Australia} \\ 
\footnotesize{$^{15}$Institut d'Astrophysique de Paris, 98 bis Boulevard Arago, F-75014 Paris, France} \\ 
\footnotesize{$^{16}$INAF-OAR, Via Frascati 33, 00040 Monte Porzio - Rome, Italy} \\ 
\footnotesize{$^{17}$Carnegie Observatories, 813 Santa Barbara Street, Pasadena, CA 91101} \\ 
\footnotesize{$^{18}$NASA/Goddard Space Flight Center, Code 662, Greenbelt, MD 20771} \\ 
\footnotesize{$^\ast$To whom correspondence should be addressed; E-mail:  pkelly@astro.berkeley.edu.}
}

\title{Multiple Images of a Highly Magnified Supernova Formed by an Early-Type Cluster Galaxy Lens\footnote{This manuscript has been accepted for publication in Science. This version has not undergone final editing. Please refer to the complete version of record at http://www.sciencemag.org/. The manuscript may not be reproduced or used in any manner that does not fall within the fair use provisions of the Copyright Act without the prior, written permission of AAAS.}}

\date{}

\begin{document}

\maketitle

\begin{sciabstract}
In 1964, Refsdal hypothesized that a supernova whose light traversed multiple paths around a strong gravitational lens could be used to measure the rate of cosmic expansion.  We report the discovery of such a system. 
In {\it Hubble Space Telescope} imaging, we have found four images of a single supernova forming an Einstein cross configuration around a redshift $z=0.54$ elliptical galaxy in the MACS\,J1149.6+2223 cluster.
The cluster's gravitational potential also creates multiple images of the $z=1.49$ spiral supernova host galaxy, and a future appearance of the supernova elsewhere in the cluster field is expected. 
The magnifications and staggered arrivals of the supernova images probe the cosmic expansion rate, as well as the distribution of matter in the galaxy and cluster lenses.
\end{sciabstract}

\newrefsegment
The possibility that the light from an exploding supernova (SN) could follow more
than a single path around an intervening strong galaxy lens to the
observer was first explored about 50 years ago \cite{refsdal64}.
Many decades of searches for SNe, however, have not identified an
explosion visible at multiple positions around a gravitational
lens.  Here we report a strongly lensed supernova
found in resolved multiple images, which we identified in the
MACS\,J1149.6+2223 \cite{ebelingedgehenry01} galaxy cluster field on 11
November 2014 (Universal Time dates are used throughout this paper).

Although the apparent positions of galaxies that are multiply imaged by a
foreground galaxy or cluster are now widely used to map the
matter distribution within the lenses, a strongly lensed background source with
a varying light curve allows distinct and powerful measurements of the
lens and cosmology, because the delay between each pair of images can be
measured.  This difference in arrival time, owing to the difference in
geometric and gravitational time delay \cite{shapiro64}, is directly
proportional to the so-called time-delay distance, and thus inversely
proportional to the Hubble constant and weakly dependent on other
cosmological parameters \cite{refsdal64,treu10,linder11,suyutreuhilbert14}. 
Conversely, for an assumed cosmological model, the time delays are a direct measurement of the
difference in gravitational potential between the multiple images and
hence greatly improve the reconstruction of the mass distribution in
the deflector \cite{Koc++06}.

After the discovery of the SBS 0957+561 A/B system 26 years ago \cite{walshcarswellweymann79}, a handful of quasi-stellar objects (quasars) multiply imaged by an intervening galaxy lens have been identified \cite{Ina++12}. Quasars strongly lensed by clusters are even more rare events, with only several known \cite{Sha++05}. The use of lensed quasars as robust probes of the distribution of matter in the lenses and of cosmology has only become possible relatively recently given the long time periods of monitoring needed to match their complex light curves \cite{koopmanstreufassnacht03,Tew++13,suyuaugerhilbert13,suyutreuhilbert14}. 
In contrast, all SNe have much simpler light curves and evolve comparatively rapidly, which makes the measurement of time delays and magnification among the multiple images substantially more straightforward.

\begin{figure*}[t]
\centering
\includegraphics[angle=0,width=6.25in]{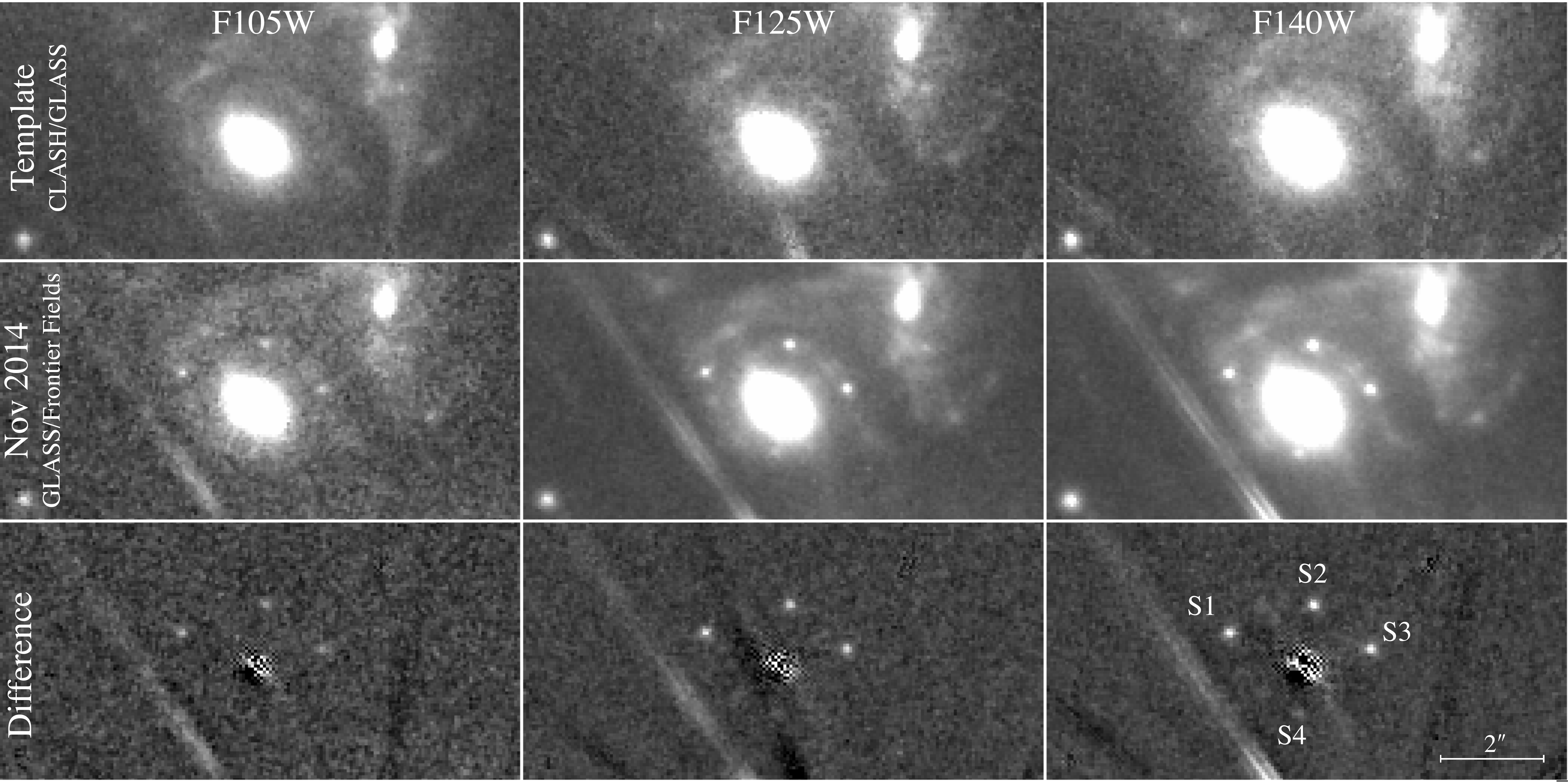}
\caption{ {\it HST} WFC3-IR images showing the simultaneous appearance
  of four point sources around a cluster member galaxy. From left to
  right the columns show imaging in the {\it F105W} filter ($Y$ band),
  {\it F125W} ($J$), and {\it F140W} ($JH$). From top to bottom the rows
  show archival imaging from the 
  Cluster Lensing And Supernova survey with Hubble (CLASH, GO-12068; PI: M.P.) program, 
  discovery epoch images from GLASS and the Hubble Frontier Fields programs, and the
  difference images.  The template images in the top row comprise
  all available archival WFC3-IR imaging in these filters, collected
  from 5 December 2010 through 10 March 2011.  The images in the middle row are the composite of all available {\it HST} imaging collected
  between 3 November and 11 November 2014 (for {\it F105W}, left),
  on 20 November 2014 ({\it F125W}, middle), and between 10 November
  and 20 November 2014 ({\it F140W}, right). The sources S1, S2, S3,
  and S4, which form an Einstein cross, are absent from all images
  obtained at earlier epochs but are clearly detected in
  the difference images along the bottom row. The line segments below
  S4 and in the lower right corner are diffraction spikes from a
  nearby bright star in the foreground.  }
\label{fig:detection}
\end{figure*}

It was recently shown that a different SN, PS1-10afx \cite{chornockbergerrest13} at redshift $z=1.38$, was strongly magnified (by a factor of $\sim30$) by an intervening galaxy at $z=1.12$ \cite{quimbywerneroguri13,quimbyoguirmore14}. The available imaging, taken from the ground, had insufficient angular resolution to separate potential multiple images of the SN, so time delays and magnifications could not be measured.
In the case presented here, the four images of the SN are
clearly resolved (Fig.~\ref{fig:detection}) with an image separation of over $2''$, thereby
presenting an ideal opportunity to carry out for the first time an
experiment similar to that suggested by Refsdal \cite{refsdal64},
leading us to name the supernova ``Refsdal.''

The Grism Lens-Amplified Survey from Space (GLASS) program [GO-13459, principal investigator (PI) T.T.] is a 140-orbit {\it Hubble Space Telescope} ({\it HST}) project that is acquiring near-infrared grism spectra of massive galaxy clusters with the primary goals of studying faint high-redshift ($z \gtrsim 6$) galaxies \cite{schmidttreubrammer14} and spatially resolved intermediate-redshift galaxies \cite{tuckerwangschmidt14}, as well as characterizing the cluster galaxy population. Wide-band near-infrared {\it F105W} and {\it F140W} exposures are taken using the Wide Field Camera 3 (WFC3) to align and calibrate the grism data, and we have been searching these images for transient sources. 

\begin{figure}[t]
\centering
\subfigure{\includegraphics[angle=0,width=6in]{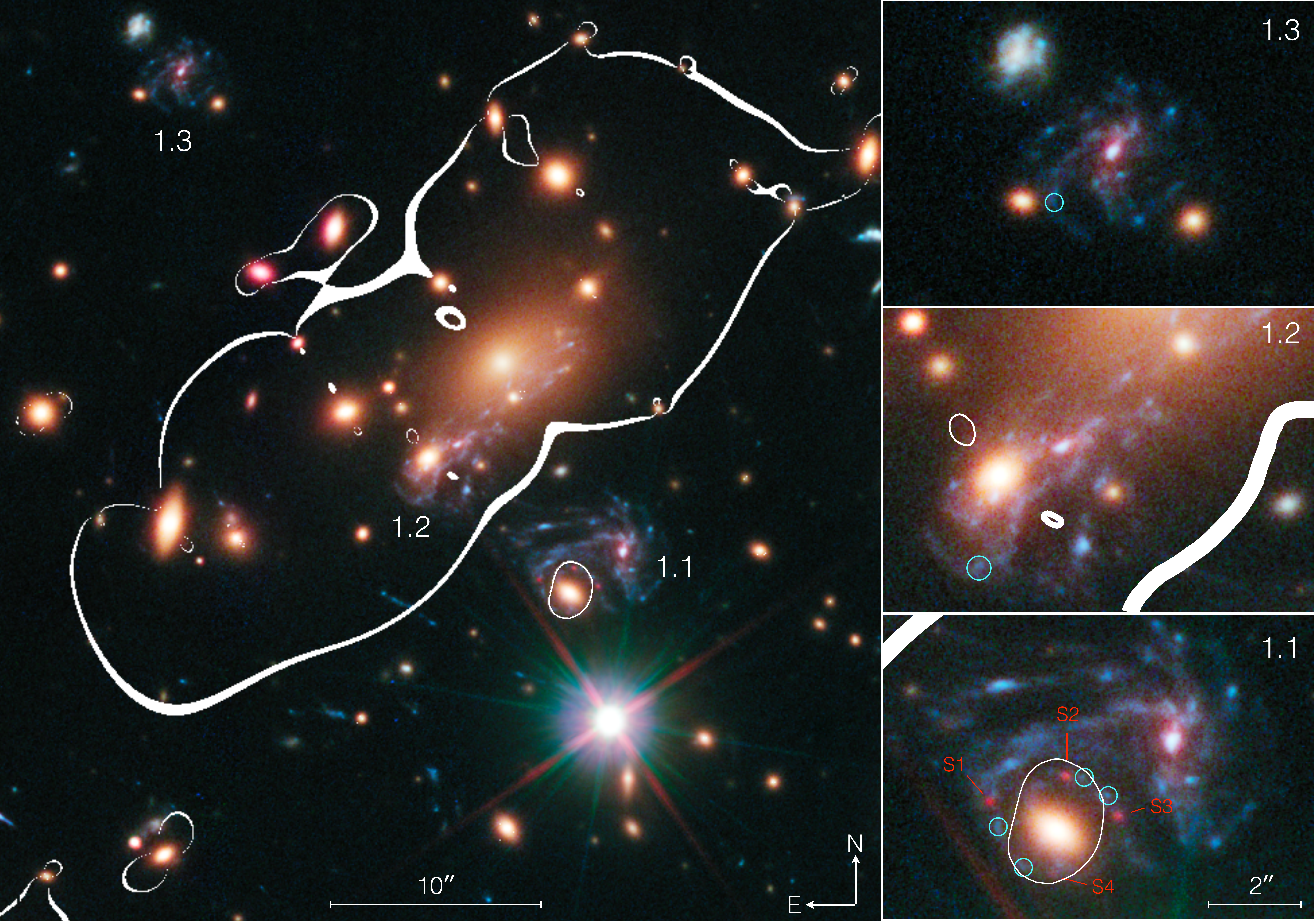}}
\caption{Color-composite image of the galaxy cluster MACSJ1149.6+2223, with critical curves for sources at the $z=1.49$ redshift of the host galaxy overlaid.  Three images of the host galaxy formed by the cluster are marked with white labels (1.1, 1.2, and 1.3) in the left panel, and each is enlarged at right. 
The four current images of SN Refsdal that we detected (labeled S1 to S4 in red) appear as red point sources in image 1.1.  Our model indicates that an image of the SN appeared in the past in image 1.3, and that one will appear in the near future in image 1.2. The extreme red hue of the SN may be somewhat exaggerated, because the blue and green channels include only data taken before the SN erupted. 
In image 1.1, both a single bright blue knot (cyan circles) and SN Refsdal are multiply imaged into four distinct locations.   
The image combines infrared and optical {\it HST} imaging data from the Frontier Fields and GLASS programs, along with images from the CLASH and the FrontierSN programs (GO-13790, PI: S.A.R.).
}
\label{fig:color}
\end{figure}

\begin{figure}[t]
\centering
\subfigure{\includegraphics[angle=0,width=6.0in]{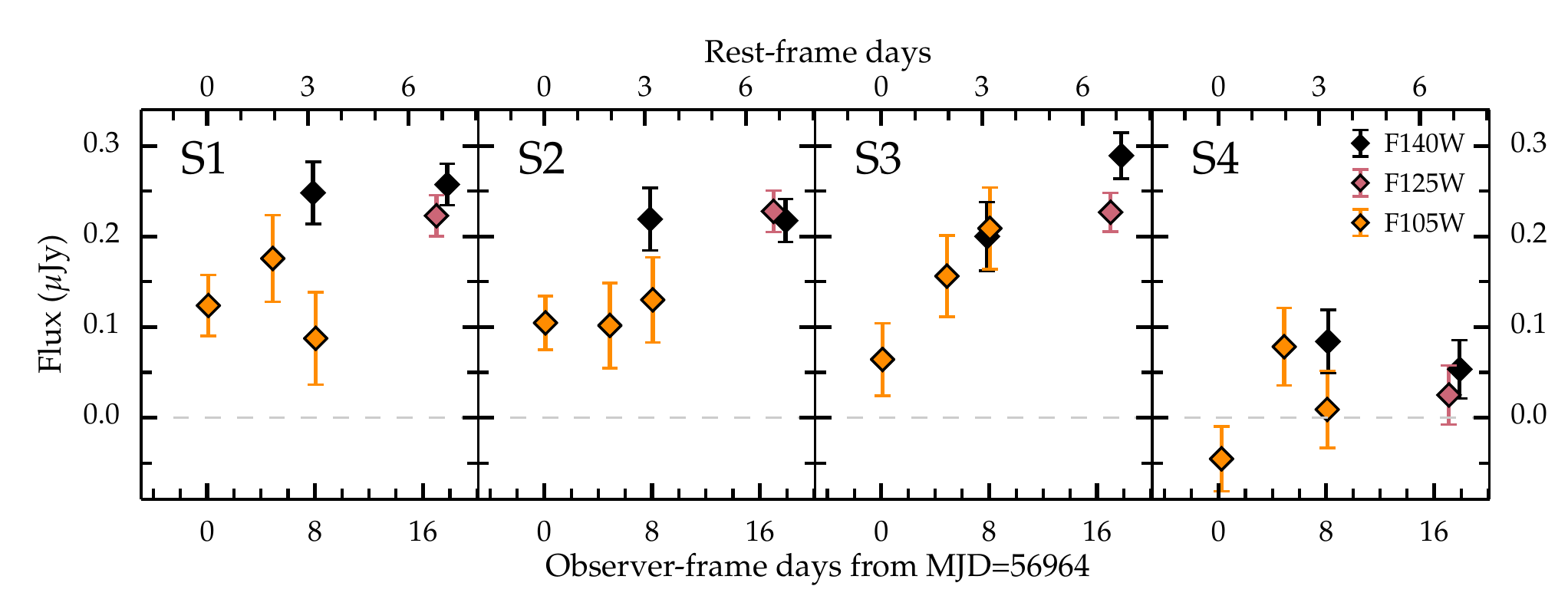}}
\caption{Light curve of the images S1 to S4 of the strongly lensed SN taken from 3 November 2014 through 20 November 2014. 
Rest-frame days assume that the SN is at the redshift of the multiply imaged spiral galaxy ($z=1.49$).
We plotted the fluxes measured in the WFC3 {\it F105W}, {\it F125W}, and {\it F140W} images of the 
MACS\,J1149.6+2223 galaxy cluster field. 
The expected time delays between images of days to weeks 
suggest that the transient must evolve over a timescale similar to that of a SN.
Our lens models generally predict that image S3 is delayed relative to S1 and S2, which is consistent with the early photometry.   
Flux uncertainties are calculated by injecting a thousand point sources into the difference images and comparing the fluxes recovered using point-spread-function fitting with the input fluxes. Error bars throughout correspond to the standard deviation of a normal distribution fitted to the histogram of the difference in flux. 
}
\label{fig:lightcurve}
\end{figure}

In the {\it F140W} GLASS images acquired on 10 November 2014, we detected the component images  
of a quadruple lens system, which we label sources S1 to S4 (Fig.~\ref{fig:detection}).
Table~\ref{tab:sources} gives the coordinates of the variable sources. 
In Figure~\ref{fig:color}, the color-composite image shows the red galaxy lens at $z=0.54$ \cite{ebelingbarrettdonovan07} surrounded by an Einstein ring formed by light from the distorted spiral host galaxy with $z=1.49$ \cite{smithebelinglimousin09}, whose nucleus is offset by $\sim3.3''$ from the center of the lensing elliptical galaxy.  
Although sources S1 and S2 do not exhibit a significant change in their fluxes during the imaging taken from 3 to 20 November 2014, the light curve of S3 is consistent with a rise in brightness during this period, which corresponds to approximately a week in the rest frame (Fig.~\ref{fig:lightcurve}; see also fig.~\ref{fig:sthreelightcurve}).  The light curve of S4 is difficult to characterize with the currently available data, because it is comparatively faint.

Based on the available data we can attempt a first preliminary classification of the SN. All known Type Ia SNe reach their peak brightness in fewer than 20 rest-frame days \cite{Ganeshalingam:2011}. 
The light curve for image S3 of SN Refsdal (see fig.~\ref{fig:sthreelightcurve}) through $>$30 days in the rest frame shows that its brightness has continued to rise for a longer period than could be expected for a SN~Ia, suggesting that it belongs to a different spectroscopic class.

Archival {\it HST} imaging and the configuration of the multiple images demonstrate that the source is not an active galactic nucleus (AGN) 
behind the galaxy and cluster lenses. 
A search of WFC3 {\it F105W}, {\it F110W}, {\it F140W}, and {\it F160W} images of MACS\,J1149.6+2223 acquired across ten separate {\it HST} visits beginning on 4 December 2010 finds no evidence for previous
variability. 
Several epochs of registered and coadded {\it F140W} imaging exhibit no significant
variation (Fig.~\ref{fig:mosaic}); seven archival epochs of {\it F160W} imaging likewise show no
significant changes. 
Evidence for previous variability would have suggested that the source is a flare from an AGN instead of a SN.
The transients detected in November 2014 are additionally several
magnitudes above the upper limits of $\sim28.5$ obtained at previous epochs (all magnitudes are in the AB system). 
Such a large increase in brightness would be very unusual for an AGN, whose light curves
typically vary at the level of a few tenths of a magnitude over several-month
time scales \cite{kaspibrandtmaoz07,bentzwalshbarth09,schmidtmarshallrix10}.  
Finally, the positions of the multiple images also constrain the redshift of the source to 1.1 to 1.7 with 95\% confidence, consistent with the $z=1.49$ redshift of the spiral galaxy lensed into the observed configuration (see fig.~\ref{fig:hostRingSED}).

The four images of SN Refsdal form an Einstein cross configuration around the massive elliptical galaxy at $z=0.54$, which adds onto and locally perturbs the cluster potential. Because the elliptical galaxy is located close to the critical lines of the cluster lens \cite{zitrinbroadhurst09},   the contribution of the galaxy cluster to the gravitational potential needs to be taken into account. As a first, simple approximation of the lensing system, we construct a single isothermal ellipsoid embedded in a strong external shear \cite{courbinchantryrevaz11}.  This yields time delays on the order of several to tens of days. S1 is generally the leading image, typically followed by S2, S3, and then S4. Magnifications are $\sim2$ for the least magnified image S4 and $\sim10$ for the other images. These magnifications, however, do not include the additional contribution from the cluster, which is expected to be very substantial, especially because earlier modeling has found a relatively flat, nearly convergent central mass distribution, which is evident from the relatively undistorted shape of the magnified spiral images \cite{zitrinbroadhurst09}. 

To account more completely for the effects of the cluster potential, we have constructed a detailed set of lens models of the entire cluster potential including the elliptical galaxy, for several different prior probability distributions and sets of constraints. These models, which are also constrained by the positions of the SN images, generally yield magnifications of $\sim$10 to 30 at the positions of the four images, and time delays on the order of days to months, in agreement with independent models \cite{Ogu14,S+J14}.  The typical arrival sequence is consistent with the predictions of the simpler, galaxy-lens model (S1, S2, and then either S3 or S4), although some models also predict different arrival orders.  These time delays are also in accord with our identification of the four newly detected sources as a multiply imaged SN, because the luminosity of a SN is not expected to vary dramatically over the time scale of less than a week in the rest frame.  

The spiral host galaxy itself is multiply imaged by the galaxy cluster \cite{zitrinbroadhurst09,smithebelinglimousin09}. Consequently, our models predict both that
the SN could be detected at future epochs in a different image of the spiral host galaxy, and that it has already appeared elsewhere in
yet another image of the spiral.  A search of archival {\it HST} imaging in both the optical ({\it F606W}, {\it F814W}, and {\it F850LP}) and infrared ({\it F105W}, {\it F125W}, {\it F140W}, and {\it F160W}) at the locations of the
multiple images of the presumed host galaxy has revealed no evidence for SN Refsdal
when these data were taken. Our set of cluster lens models predicts that the SN will appear in
the central image of the spiral host galaxy, at an approximate position of $\alpha=11^{\rm h}49^{\rm m}36.01^{\rm s}$, $\delta=+22^{\circ}23'48.13''$ (J2000.0) at a future time, within a year to a decade from now (2015 to 2025). This is in broad agreement with independent model predictions \cite{Ogu14,S+J14}. The uncertainties highlight the power of a time-delay measurement to constrain lens models.

The archival {\it HST} imaging and the configuration show that this is a multiply imaged SN. This discovery demonstrates in principle the feasibility of the experiment suggested five decades ago by Refsdal \cite{refsdal64}, consisting of using the time delays between the multiple images of the SN to constrain the foreground mass distribution, and eventually the geometry and content of the universe. 

\renewcommand\refname{References and Notes}

\printbibliography

\paragraph*{Acknowledgements} 
This work is based on data obtained with the NASA/ESA {\it Hubble Space
Telescope}.  We thank Ori Fox, WeiKang Zheng, Josh
Bloom, Charles Keeton, Jon Mauerhan, Chuck Steidel, and Allison Strom 
for helpful discussions, as well as the Space Telescope Science Institute
(STScI) and Director Matt Mountain for supporting our proposal for follow-up
observations. GLASS is supported by NASA through {\it HST} grant GO-13459.
Support for SAR was provided by NASA through Hubble Fellowship grant HST-HF-51312.01 awarded by STScI, which is operated by the Association of Universities for Research in Astronomy for NASA, under contract NAS 5-26555.  Follow-up imaging through the FrontierSN program is supported by NASA through {\it HST} grant GO-13386. AVF's group at the University of California Berkeley has received generous financial
assistance from the Christopher R. Redlich Fund, the TABASGO
Foundation, Gary and Cynthia Bengier, and NSF grant AST-1211916. The Dark Cosmology Centre 
is funded by the Danish National Research Foundation. Support for AZ was provided by NASA through Hubble Fellowship grant HF2-51334.001-A awarded by STScI. Supernova research at Rutgers University is supported in part by NSF CAREER award AST-0847157 to SWJ. JCM is supported by the NSF grant AST-1313484. RG acknowledges the CNES for financial support on the GLASS project.
Some of the data presented herein were obtained at the W. M. Keck
Observatory, which is operated as a scientific partnership among the
California Institute of Technology, the University of California, and
NASA; the observatory was made possible by the generous financial
support of the W. M. Keck Foundation.
The {\it HST} imaging data used in this paper can be obtained from the Barbara A. Mikulski Archive for Space Telescopes at \url{https://archive.stsci.edu},
while the Keck-I LRIS spectra can be obtained at \url{http://hercules.berkeley.edu/database/}.

\begin{table*}
\centering
\begin{tabular}{cccc}
\hline
Name & $\alpha$ (J2000) & $\delta$ (J2000) \\
\hline
S1 & 11$^{\rm h}$49$^{\rm m}$35.574$^{\rm s}$ & +22$^{\circ}$23$'$44.26$''$ \\
S2 & 11$^{\rm h}$49$^{\rm m}$35.451$^{\rm s}$ & +22$^{\circ}$23$'$44.84$''$ \\
S3 & 11$^{\rm h}$49$^{\rm m}$35.369$^{\rm s}$ & +22$^{\circ}$23$'$43.95$''$ \\
S4 & 11$^{\rm h}$49$^{\rm m}$35.472$^{\rm s}$ & +22$^{\circ}$23$'$42.62$''$ \\
\hline
\end{tabular}
\caption{Coordinates of the transient point sources detected around the cluster galaxy lens, in J2000 right ascension and declination.}
\label{tab:sources}
\end{table*}

\renewcommand{\thetable}{S\arabic{table}}
\renewcommand{\thefigure}{S\arabic{figure}}
\setcounter{figure}{0} 
\setcounter{table}{0} 

\clearpage
\begin{center}
\section*{\Large Supplementary Materials}
\end{center}

\paragraph{This PDF file includes the following:} 

\newenvironment{myitemize}
{ \begin{itemize}
    \setlength{\itemsep}{0pt}
    \setlength{\parskip}{0pt}
    \setlength{\parsep}{0pt}     }
{ \end{itemize}             }

\begin{myitemize}
\itemsep0em 
\setlength{\itemindent}{4.5pt}
\item[] Material and Methods
\item[] Supplementary Text
\item[] Figures S1 to S4
\item[] Tables S1 to S2
\item[] References 29--34
\end{myitemize}

\paragraph*{\large Materials and Methods}

\subparagraph{Modeling of the Galaxy Cluster and Elliptical Galaxy Lensing System.}

For this paper, we constructed a set of cluster lens models based on a method which relies on the approximation that light traces mass up to a smoothing factor, for both the luminous and dark matter \cite{Zit++09}. This set of models spans different priors and sets of constraints. We used as positional constraints all the secure systems commonly agreed upon in the literature \cite[e.g.,][]{Z+B09,Ogu14,S+J14}.
We also include the four positions of the SN as additional constraints. 
Uncertainty in the predictions for the time delays and magnifications arises from the complexity of the system and residual degeneracies inherent to lensing reconstructions. However, we can draw a few general conclusions. Our set of models (see example in Figure~\ref{fig:color}) yields typical magnification factors of 10--30 for the four SN images, although magnifications up to $\sim100$ can be reached in some extreme cases. The predicted time delays range between a few days and a few months, and up to $\sim$1 year in some models. Correspondingly, the arrival order is quite uncertain even though typically S1 and S2 arrive before S3 and S4.

\paragraph*{\large Supplementary Text}

\subparagraph{Previous and Future Appearances of SN Refsdal in Multiple Images of its Host Galaxy.}

The lens model predicts that SN Refsdal might have appeared some $\sim$20 years ago in image 1.3 \cite[labeled 1.1 by][]{Z+B09}. The SN may additionally appear in image 1.2 \cite{zitrinbroadhurst09,rauvegettiwhite14} within the next decade (2015--2025). The model also suggests that SN Refsdal could have appeared some $\sim40$--50 years ago in the partial spiral image 1.4 \cite{zitrinbroadhurst09}, although the SN position is believed not to be present in this multiple image \cite{rauvegettiwhite14}. These estimates are highly uncertain, because time delays on these cluster scales depend very sensitively on the exact expected position, and a fraction of the spiral images are only partial \cite{zitrinfabrismerten14}. 

\subparagraph{Observed Light Curve of Image S3.}

In order to test whether the properties of SN Refsdal are consistent those of Type Ia SNe, we show in Figure~\ref{fig:sthreelightcurve} a light curve for image S3 consisting of all of the available data at the time of writing. The expected light curves for SN Ia (scaled to fit the data as well as possible) are shown as a grey shaded area. The best-fitting SN Ia template provides a very poor fit to the data, with $\chi^2=181$ for 28 degrees of freedom. Thus, we conclude that SN Refsdal is very unlikely to be a SN~Ia, unless it is extremely peculiar, or the light curve is affected by extreme microlensing \cite{D+K06}. Additional data are needed to eliminate these possibilities.

\subparagraph{Absence of Variability in Prior Imaging.}
To check for past episodes of variability from this source, we performed a visual search of difference images constructed from archival {\it HST} imaging in both the optical ({\it F435W, F606W, F814W,} and {\it F850LP}) and infrared ({\it F105W, F125W, F140W,} and {\it F160W}).  We found no evidence of any transient sources prior to 3 November 2014 at the S1--S4 positions of the images of SN Refsdal. 

\subparagraph{Association of the SN with the Strongly Distorted Spiral Galaxy.}
Figure~\ref{fig:hostRingSED} shows that the broad-band spectral energy distribution (SED) 
of the Einstein ring that surrounds the four images of the SN matches that of the $z=1.49$ spiral galaxy, after subtracting contaminating light from the underlying elliptical-galaxy lens. 
The subtraction is accomplished by scaling the SED of the early-type galaxy lens until its {\it F140W} 
flux matches that measured for the Einstein ring, and removing the scaled SED.

\begin{figure*}
\centering
\includegraphics[angle=0,width=6.25in]{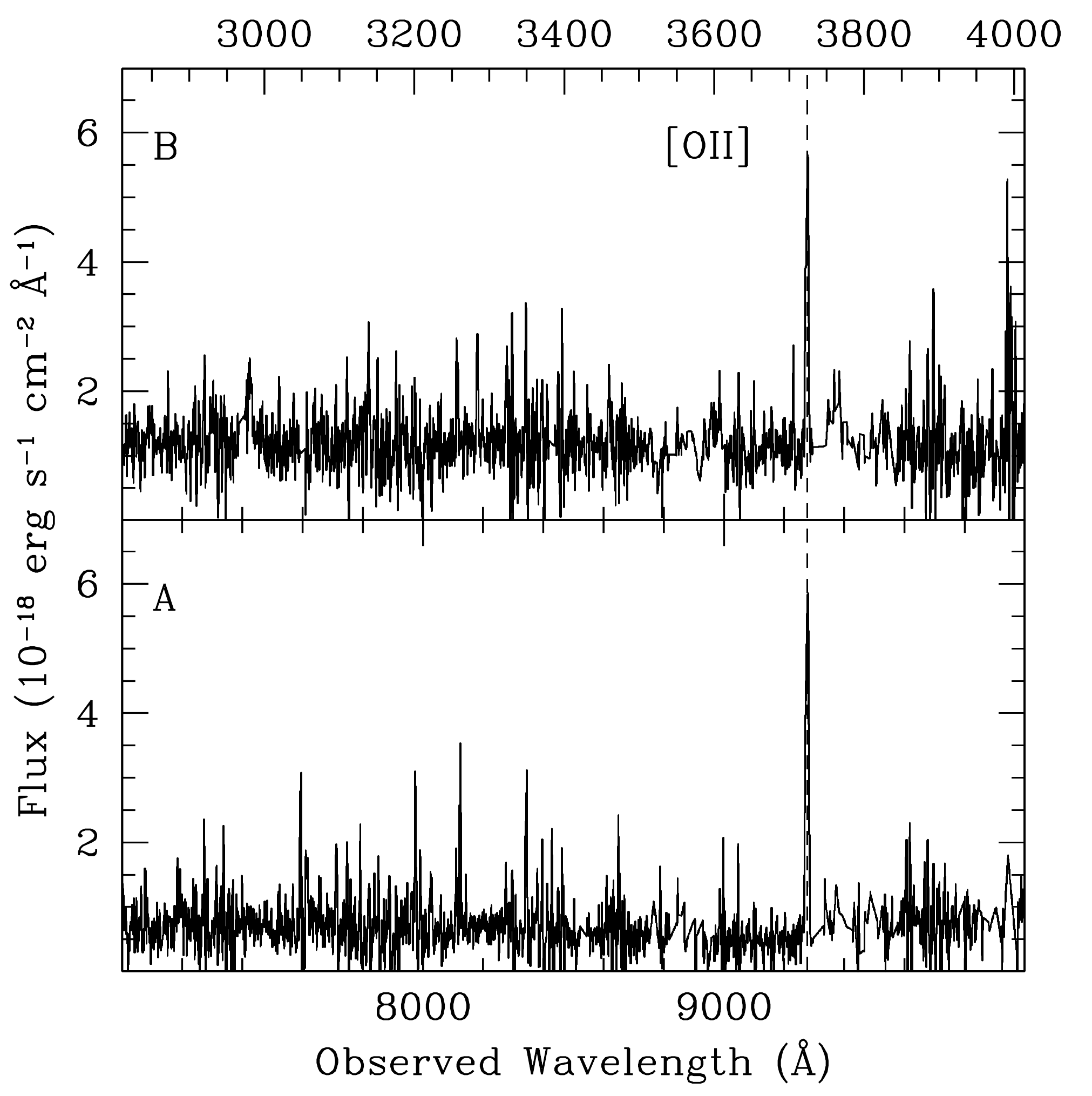}
\caption{ A spectrum of the SN Refsdal host galaxy, taken with the Keck telescope's Low Resolution Imaging Spectrometer \cite[][LRIS]{okecohen95} on 20 November 2014.  For these observations the LRIS long slit was aligned along a portion of the Einstein ring, between the positions of sources S1 and S2.  The [O II] $\lambda$3727 emission line is clearly detected at an observer-frame wavelength of 9277 \AA, confirming the redshift of the Einstein ring at $z=1.49$.  The top axis marks the rest-frame wavelength in Angstroms for this redshift solution. }
\label{fig:hostSpec}
\end{figure*}

In Figure~\ref{fig:hostSpec} we show a spectrum from the location of
the Einstein ring, collected with the Keck-I telescope in Hawaii on 
20 November 2014.  An emission feature at 9277 \AA\ is observed from
knots in the Einstein ring that are distinct from the SN source
positions, but spatially associated with the same ring feature.  We identify
this as [\ionpat{O}{ii}] $\lambda$3727, giving further
confirmation that the ring around the lensing galaxy is at $z=1.49$,
the same redshift as the presumed spiral-galaxy host behind the
cluster. No emission from SN Refsdal is apparent in this spectrum.

\subparagraph{Photometry of Multiple Images of the SN.}

In Table~\ref{tab:photometry}, the magnitudes of the sources are given 
for the three {\it F105W}, single {\it F125W}, and two {\it F140W} visits, measured using point-spread-function fitting
photometry.

\begin{figure*}
\centering
\includegraphics[angle=0,width=6.25in]{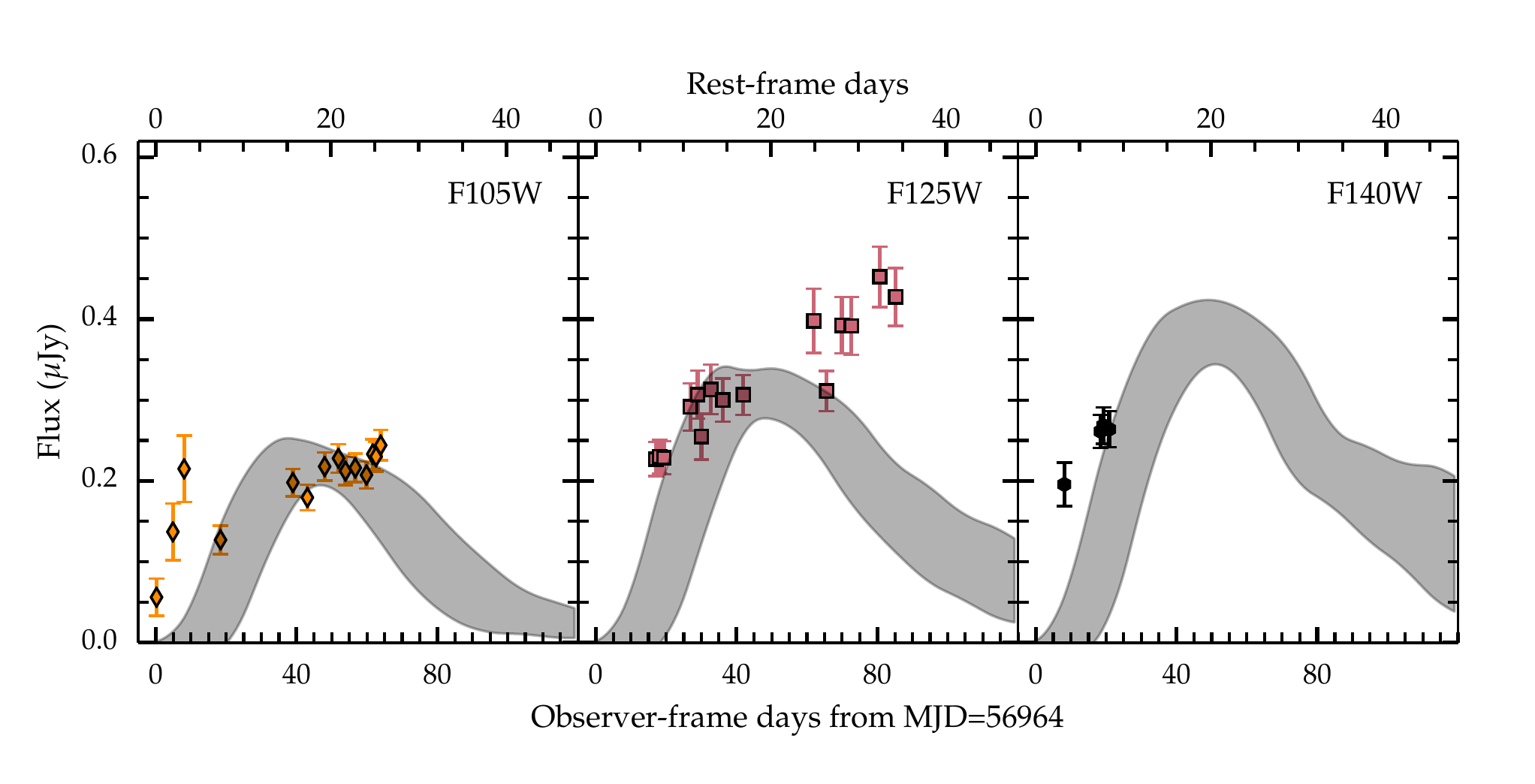}
\caption{ Light curve of image S3 of SN Refsdal through January 2015.  
The SN shows a steady rise in brightness over a period of more than 80 days in the observer frame ($>$30 days in the rest frame).
A normal Type Ia SN (shown as a gray shaded region) would have reached peak brightness and begun to fade during the time period. }
\label{fig:sthreelightcurve}
\end{figure*}

\begin{figure*}
\centering
\includegraphics[angle=0,width=6.25in]{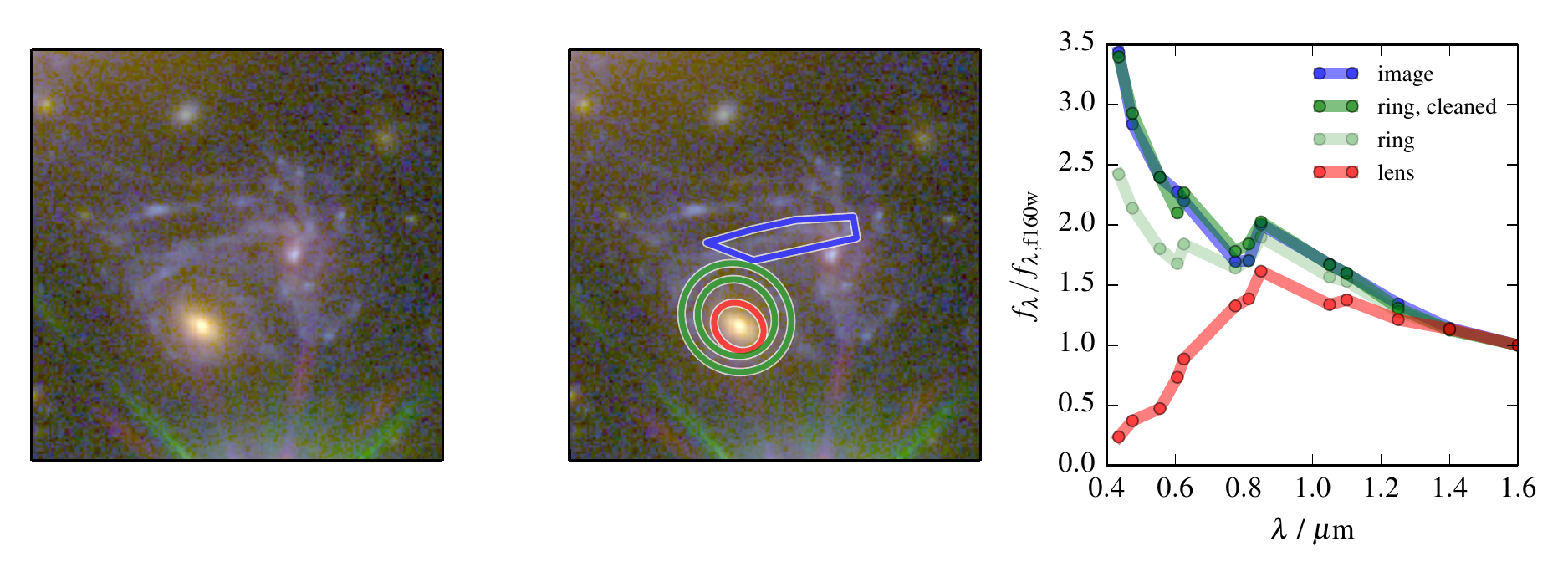}
\caption{ SEDs of the $z=1.49$ spiral host galaxy, Einstein ring, and elliptical-galaxy lens measured from pre-explosion {\it HST} imaging.  After subtracting the underlying light from the red elliptical galaxy lens, the SED of the Einstein ring matches that of the strongly distorted $z=1.49$ spiral galaxy. 
This indicates that the SN, whose images are embedded in the Einstein ring, is spatially coincident with the spiral galaxy.  Lens modeling also strongly suggests that the ring is associated with the main lensed galaxy \cite{zitrinbroadhurst09}. The blue polygon is an aperture enclosing the spiral host galaxy light, the green elliptical annulus aperture encloses the Einstein ring where the multiple SN images appeared, and the red elliptical aperture encloses the lens.  All of the SEDs are normalized to the {\it F160W} flux. }
\label{fig:hostRingSED}
\end{figure*} 

\begin{figure*}
\centering
\subfigure{\includegraphics[angle=0,width=3.0in]{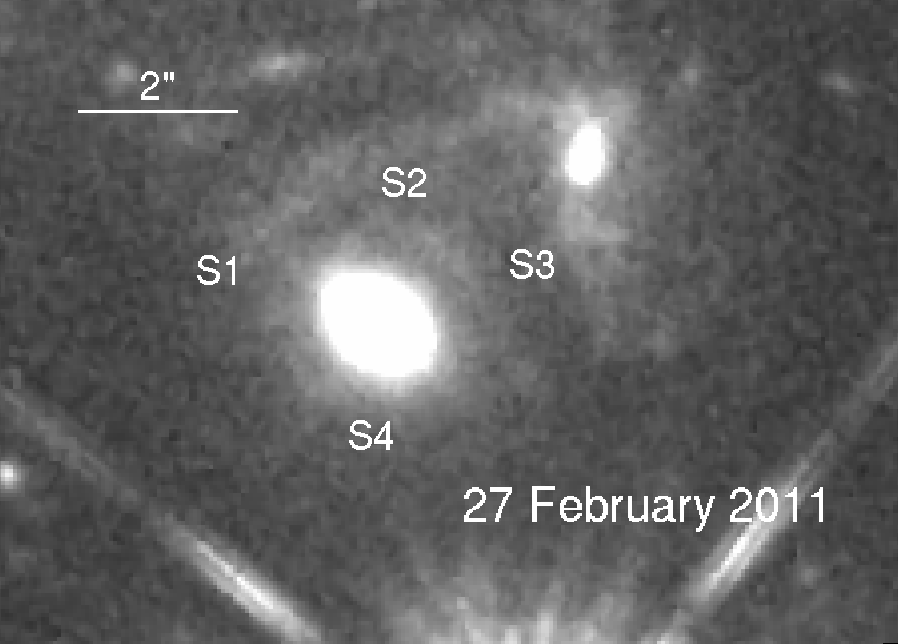}
\includegraphics[angle=0,width=3.0in]{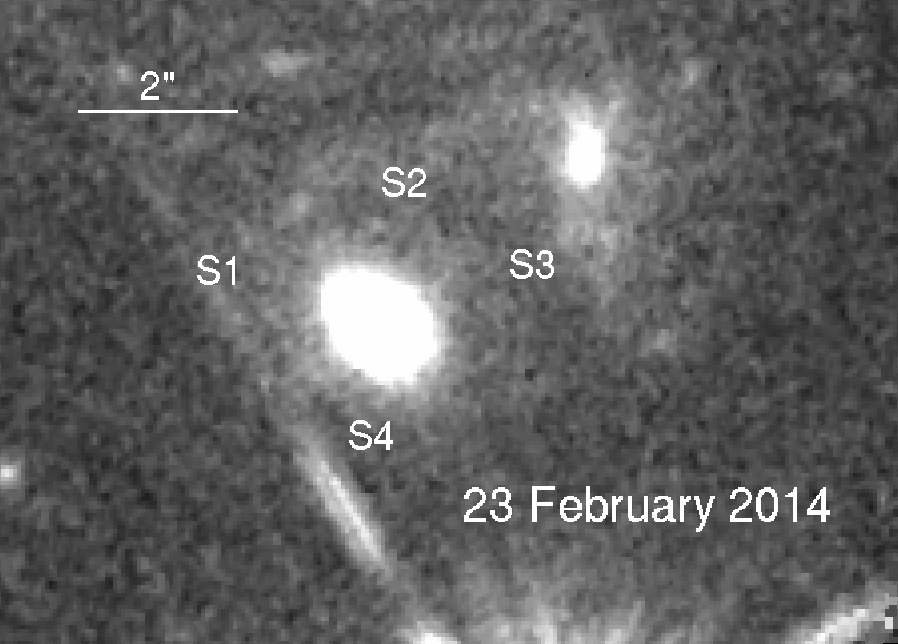}}
\subfigure{\includegraphics[angle=0,width=3.0in]{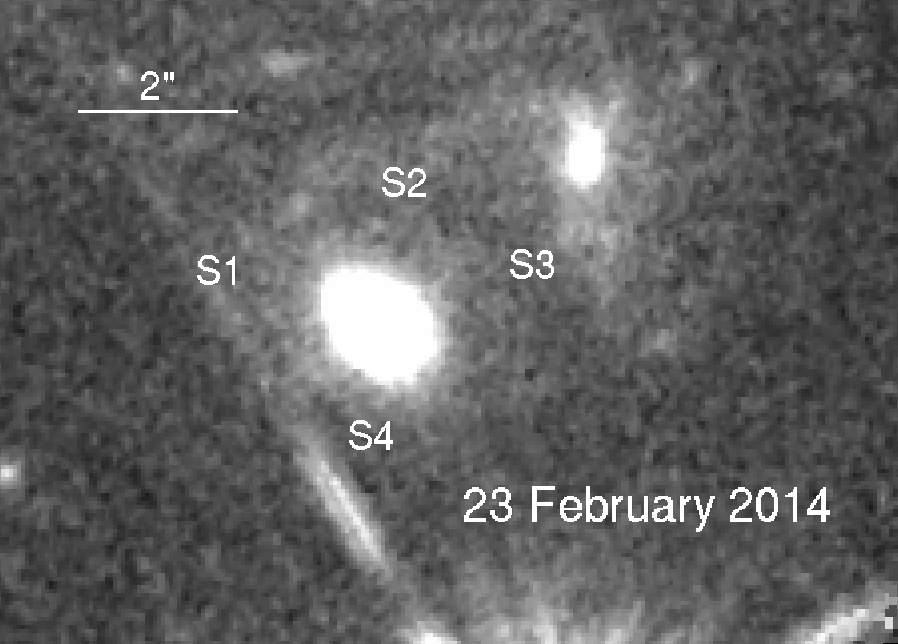}
\includegraphics[angle=0,width=3.0in]{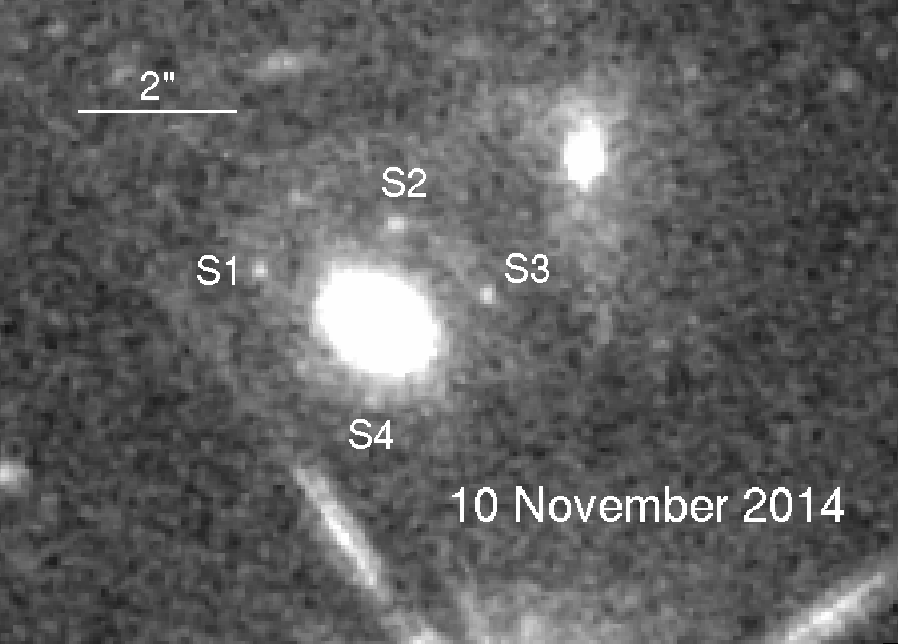}}
\caption{Images of the lensing system from archival {\it HST} WFC3-IR observations in the {\it F140W} filter. All exposures obtained prior to 3 November 2014 show no evidence for variability at any of the positions associated with SN Refsdal.}
\label{fig:mosaic}
\end{figure*} 

\begin{table}
\centering
\begin{tabular}{ccccc}
\hline
Date & MJD & Filter & Exposure Time (s) & {\it HST} Program ID \\
\hline
2004 Apr 22.7 & 53117.7 &  {\it F814W}  &  4590 &  9722 \\
2006 May 25.5 & 53880.5 &  {\it F814W}  &  2184 & 10493 \\
2010 Dec 04.9 & 55534.9 &  {\it F850LP} &  1032 & 12068 \\
2010 Dec 05.0 & 55535.0 &  {\it F125W}  &  1509 & 12068 \\
2010 Dec 05.0 & 55535.0 &  {\it F160W}  &  1006 & 12068 \\
2011 Jan 16.0 & 55577.0 &  {\it F110W}  &  1509 & 12068 \\
2011 Jan 16.0 & 55577.0 &  {\it F606W}  &  1032 & 12068 \\
2011 Jan 16.0 & 55577.0 &  {\it F850LP} &  1044 & 12068 \\
2011 Jan 16.1 & 55577.1 &  {\it F160W}  &  1006 & 12068 \\
2011 Jan 30.6 & 55591.6 &  {\it F850LP} &  1032 & 12068 \\
2011 Jan 30.7 & 55591.7 &  {\it F105W}  &  1509 & 12068 \\
2011 Jan 30.7 & 55591.7 &  {\it F140W}  &  1006 & 12068 \\
2011 Feb 13.3 & 55605.3 &  {\it F625W}  &  1015 & 12068 \\
2011 Feb 27.0 & 55619.0 &  {\it F850LP} &  1032 & 12068 \\
2011 Feb 27.1 & 55619.1 &  {\it F775W}  &   994 & 12068 \\
2011 Feb 27.1 & 55619.1 &  {\it F160W}  &  1509 & 12068 \\
2011 Feb 27.1 & 55619.1 &  {\it F110W}  &   906 & 12068 \\
2011 Feb 27.5 & 55619.5 &  {\it F606W}  &  1032 & 12068 \\
2011 Feb 27.6 & 55619.6 &  {\it F140W}  &  1306 & 12068 \\
2011 Feb 27.7 & 55619.7 &  {\it F105W}  &  1306 & 12068 \\
2011 Mar 09.8 & 55629.8 &  {\it F160W}  &  1509 & 12068 \\
2011 Mar 09.8 & 55629.8 &  {\it F625W}  &  1032 & 12068 \\
2011 Mar 09.8 & 55629.8 &  {\it F775W}  &  1053 & 12068 \\
2011 Mar 09.9 & 55629.9 &  {\it F125W}  &  1006 & 12068 \\
2013 Nov 02.1 & 56598.1 &  {\it F160W}  &  5508 & 13504 \\
2014 Feb 23.4 & 56711.4 &  {\it F105W}  &   762 & 13459 \\
2014 Feb 23.7 & 56711.7 &  {\it F140W}  &   762 & 13459 \\
2014 Feb 23.9 & 56711.9 &  {\it F105W}  &   406 & 13459 \\
2014 Feb 25.6 & 56713.6 &  {\it F105W}  &   762 & 13459 \\
2014 Apr 14.7 & 56761.7 &  {\it F814W}  &  5246 & 13504 \\
\hline
\end{tabular}
\caption{Past epochs of {\it HST} imaging on the MACS J1149.6+2223 field.  In a visual inspection of difference images constructed from these data, no evidence for any transient source in any image of the SN Refsdal host galaxy was found. }
\label{tab:allepochs}
\end{table}

\begin{table}
\centering
\begin{tabular}{crrrrrrr} 
\hline
Source & Date & MJD & Filter & Flux ($\mu$Jy) & Flux $\sigma$ & AB Mag & Mag $\sigma$ \\
\hline
S1 & 2014 Nov 10.9 &  56971.9 & {\it F140W}  &  0.248 &  0.034 &  25.41 &    0.15  \\
S1 & 2014 Nov 20.8 &  56981.8 & {\it F140W}  &  0.257 &  0.023 &  25.38 &    0.10  \\
S1 & 2014 Nov 20.0 &  56981.0 & {\it F125W}  &  0.223 &  0.023 &  25.53 &    0.11  \\
S1 & 2014 Nov 03.1 &  56964.1 & {\it F105W}  &  0.124 &  0.034 &  26.16 &    0.30  \\
S1 & 2014 Nov 07.9 &  56968.9 & {\it F105W}  &  0.176 &  0.048 &  25.79 &    0.30  \\
S1 & 2014 Nov 11.1 &  56972.1 & {\it F105W}  &  0.087 &  0.051 &  26.46 &    0.64  \\
S2 & 2014 Nov 10.9 &  56971.9 & {\it F140W}  &  0.219 &  0.034 &  25.54 &    0.17  \\
S2 & 2014 Nov 20.9 &  56981.9 & {\it F140W}  &  0.217 &  0.024 &  25.55 &    0.12  \\
S2 & 2014 Nov 20.0 &  56981.0 & {\it F125W}  &  0.228 &  0.023 &  25.50 &    0.11  \\
S2 & 2014 Nov 03.1 &  56964.1 & {\it F105W}  &  0.105 &  0.030 &  26.33 &    0.31  \\
S2 & 2014 Nov 07.9 &  56968.9 & {\it F105W}  &  0.102 &  0.047 &  26.44 &    0.50  \\
S2 & 2014 Nov 11.1 &  56972.1 & {\it F105W}  &  0.130 &  0.047 &  26.03 &    0.39  \\
S3 & 2014 Nov 10.9 &  56971.9 & {\it F140W}  &  0.200 &  0.038 &  25.65 &    0.21  \\
S3 & 2014 Nov 20.8 &  56981.8 & {\it F140W}  &  0.289 &  0.025 &  25.26 &    0.09  \\
S3 & 2014 Nov 20.0 &  56981.0 & {\it F125W}  &  0.227 &  0.021 &  25.51 &    0.10  \\
S3 & 2014 Nov 03.1 &  56964.1 & {\it F105W}  &  0.064 &  0.040 &  26.93 &    0.68  \\
S3 & 2014 Nov 07.9 &  56968.9 & {\it F105W}  &  0.156 &  0.045 &  25.90 &    0.31  \\
S3 & 2014 Nov 11.1 &  56972.1 & {\it F105W}  &  0.209 &  0.045 &  25.67 &    0.23  \\
S4 & 2014 Nov 11.1 &  56972.1 & {\it F140W}  &  0.084 &  0.035 &  26.59 &    0.45  \\
S4 & 2014 Nov 20.9 &  56981.9 & {\it F140W}  &  0.053 &  0.032 &  27.08 &    0.66  \\
S4 & 2014 Nov 20.1 &  56981.1 & {\it F125W}  &  0.025 &  0.033 &  27.90 &    1.43  \\
S4 & 2014 Nov 03.2 &  56964.2 & {\it F105W}  & -0.045 &  0.036 & $>$26.70 & \nodata \\
S4 & 2014 Nov 07.9 &  56968.9 & {\it F105W}  &  0.078 &  0.043 &  26.66 &    0.60  \\
S4 & 2014 Nov 11.1 &  56972.1 & {\it F105W}  &  0.009 &  0.042 &  29.01 &    5.07  \\
\hline
\end{tabular}
\caption{{\it HST} WFC3-IR photometry. The observed fluxes were
  measured with point-spread-function fitting on difference images
  using template images constructed from all WFC3-IR images collected
  prior to November 2014. In cases where the source is not detected
  and noise from the sky results in a measurement of negative flux, we
  report the magnitude as a $3\sigma$ upper limit.}
\label{tab:photometry}
\end{table}

\end{document}